\documentclass[twocolumn, 
prb,preprintnumbers,amsmath,amssymb,floatfix]{revtex4}

\usepackage{graphicx}
\usepackage{dcolumn}

\begin{document}

\title{Electron-Hole Symmetry and Magnetic Coupling \\
  in Antiferromagnetic LaOFeAs}
\author{Z. P. Yin$^1$, S. Leb\`egue$^{1,2}$, M. J. Han$^1$, B. Neal$^1$, S. Y. Savrasov$^1$, 
   and W. E. Pickett$^1$}
\affiliation{$^1$Department of Physics, University of California Davis, 
  Davis, CA 95616}
\affiliation{
 $^2$Laboratoire de Cristallographie et de Mod\'elisation des Mat\'eriaux
 Min\'eraux et Biologiques,
 UMR 7036, CNRS-Universit\'e Henri Poincar\'e, B.P. 239, F-54506
 Vandoeuvre-l\`es-Nancy, France
}
\date{\today}
\begin{abstract}
When either electron or hole doped at concentrations $x\sim 0.1$,
the LaOFeAs family displays remarkably high temperature superconductivity 
with T$_c$ up to 55 K.  In the most energetically stable
$\vec Q_M = (\pi,\pi)$ antiferromagnetic (AFM) phase comprised of
tetragonal-symmetry breaking alternating chains of aligned spins,  
there is a deep pseudogap in the Fe $3d$ states centered at the Fermi
energy, and very strong magnetophonon coupling is uncovered.  
Doping (of either sign)
beyond $x \sim 0.1$ results in  
Fe $3d$ heavy mass carriers ($m^*\sim 4-8$) with a large Fermi surface. 
Calculated Fe-Fe transverse exchange couplings $J_{ij}(R)$ 
reveal that exchange coupling is strongly
dependent on the AFM symmetry and Fe-As distance.
\end{abstract}
\maketitle

Since the appearance of copper oxide high temperature superconductors
(HTS) two decades ago,\cite{hts} there has been a determined but underfunded effort
to discover related superconductors in two-dimensional (2D) transition
metal oxides (TMO), borides, nitrides, etc.  Promising developments in this
area include Li$_x$NbO$_2$,[\onlinecite{linbo2}], 
Sr$_2$RuO$_4$,[\onlinecite{sr2ruo4}] Na$_x$CoO$_2$,[\onlinecite{naxcoo2}],
and Cu$_x$TiSe$_2$,[\onlinecite{cutise2}] but all have superconducting 
critical temperature
T$_c$ of 5K or less.  The most striking discovery was that of 
electron-doped hafnium nitride semiconductor (HfNCl) [\onlinecite{hfncl}] 
with T$_c$ = 25 K.  The other distinctive breakthrough,\cite{mgb2} MgB$_2$ (T$_c$=40 K),
has strong 2D features but contains only $s,p$ elements. Recently, design
of possible TMO superconductors has been stimulated by a specific
approach outlined by Chaloupka and Khaluillin.\cite{giniyat}

The simmering state of superconductor discovery has been
re-ignited by discovery of
a new class of 
layered transition metal pnictides
${\cal R}$O${\cal T}$Pn, where ${\cal R}$ is a trivalent rare earth ion,
${\cal T}$ is a late transition metal ion, and Pn is a pnictogen atom.
The breakthrough of T$_c$=26 K (T$^{onset}_c$=32 K) was 
reported\cite{1stFeAs} for
$0.04 \leq x \leq 0.12$ electron doped LaO$_{1-x}$F$_x$FeAs, followed by the demonstration
that hole-doping\cite{hole-doped} in 
La$_{1-x}$Sr$_x$OFeAs, $0.09 \leq x \leq 0.20$, leads to a similar
value of T$_c$.  These values of T$_c$ have now been superseded by the
finding that replacement of La by Ce,\cite{CeFeAs} Pr,\cite{PrFeAs} 
Nd,\cite{NdFeAs} Sm,\cite{SmFeAs,SmFeAs2} and Gd\cite{GdFeAs} result in 
T$_c$ = 41-55 K,
substantially higher than in any materials except for the cuprate HTS. 

\begin{figure}[tbp]
\rotatebox{-00}
{\resizebox{5.0cm}{6.0cm}{\includegraphics{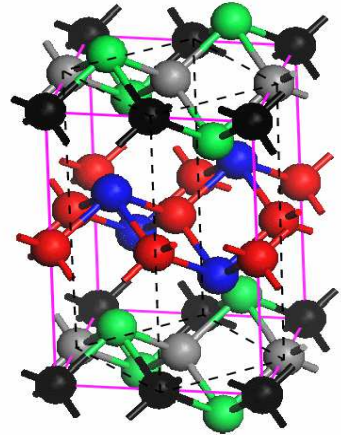}}}
\caption{(color online) The $Q_M$ antiferromagnetic structure of LaOFeAs,
with different shades of Fe
atoms (top and bottoms planes) denoting the opposing directions of spins 
in the $Q_M$ AFM phase. Fe atoms lie
on a square sublattice coordinated tetrahedrally by As atoms, separated
by LaO layers (center of figure) of similar structure. The dashed 
lines indicate the nonmagnetic primitive cell.}
\label{structure}
\end{figure}

The transport, magnetic, and superconducting properties of 
LaO$_{1-x}$F$_x$FeAs depend
strongly on doping.\cite{1stFeAs,hole-doped,ORNL}  
Most interestingly, a kink is observed\cite{CeFeAs} in the
resistivity of the stoichiometric (``undoped'' but conducting) 
compound, which has been identified with the
onset of antiferromagnetism (AFM).  As a result, the original focus on the
nonmagnetic LaOFeAs compound  
switched to an AFM ground state, in which
the two Fe atoms in the primitive cell have oppositely oriented moments.
Due to the structure of the FeAs layer, shown in Fig.
\ref{structure}, that requires two Fe
atoms in the primitive cell, this ordering represents a $Q$=0 AFM state.

The basic electronic structure of this
class of compounds was presented for LaOFeP, superconducting at 
5 K [\onlinecite{LaOFeP}], by Leb\`egue.\cite{lebegue}
The electronic structure of paramagnetic LaOFeAs is similar,
and its (actual or incipient) magnetic instabilities have been described 
by Singh and Du,\cite{SinghDu} who found that the Fermi level (E$_F$) lies on
the edge of a peak in the density of states (DOS), making the electronic
structure strongly electron-hole {\it asymmetric}.  The Fermi
surfaces are dominated by zone center and zone corner cylinders, which 
underlie several
models of both magnetic\cite{interband.deg} 
and superconducting.\cite{mazin,disconnectFS,LsingletStriplet,generic2band} properties.
Cao {\it et al.}\cite{hirschfeld} and Ma and 
Lu\cite{MaLu-afm} demonstrated that a $Q$=0 AFM state (mentioned above)
is energetically favored,  
but coincidentally (because the electronic structure is substantially
different) still leaves E$_F$ on the edge of a DOS peak, {\it i.e.} strongly
particle-hole asymmetric.  
In both paramagnetic and $Q=0$ AFM states a degenerate
$d_{xz}, d_{yz}$ pair of Fe orbitals remains roughly half-filled,
suggesting possible spontaneous symmetry breaking to eliminate
the degeneracy.\cite{interband.deg,orb-deg}  Such degeneracies have 
attracted attention in transition metal oxides.\cite{srfeo2}

Subsequently
it was reported by Dong {\it et al.}\cite{Qo-afm} that a $\vec Q_M =
(\pi,\pi,0)$ $\sqrt{2} \times \sqrt{2}$ AFM state lies 
substantially lower still in energy. 
The spin arrangement consists of Fe chains
of aligned spins along one direction (which we take to be the $x$-axis) of the
square Fe sublattice, with alternate chains having opposite spin direction.
This $\vec Q_M$ ordering is what might be
expected from the (approximate) nesting of Fermi surfaces in the primitive
cell, but the calculated moments are large (1.72 $\mu_B$ in the $Q$=0 phase,
1.87 $\mu_B$ for $Q_M$) 
and thus is far removed from a
`spin density wave' description. Neutron scattering\cite{NIST,ORNL2} and
x-ray scattering\cite{ORNL2} have confirmed this in-plane ordering, and
reveal that alternating planes of Fe spins are antialigned, {\it i.e.}
the true ordering is $(\pi,\pi,\pi)$.

\begin{figure}[tbp]
\rotatebox{-90}
{\resizebox{7.0cm}{8.0cm}{\includegraphics{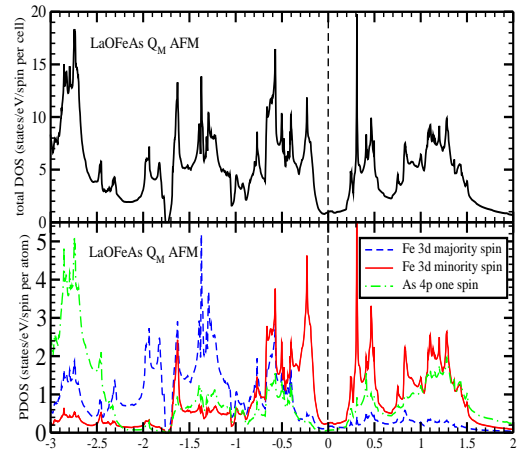}}}
\caption{(color online) Top panel: total DOS for the $Q_M$ AFM phase.
Bottom panel: spin resolved Fe $3d$ DOS, showing majority filled and
minority half-filled up to the pseudogap, 
and the As $4p$ DOS.}
\label{dos}
\end{figure}

\begin{figure}[tbp]
\rotatebox{-90}
{\resizebox{5.0cm}{7.0cm}{\includegraphics{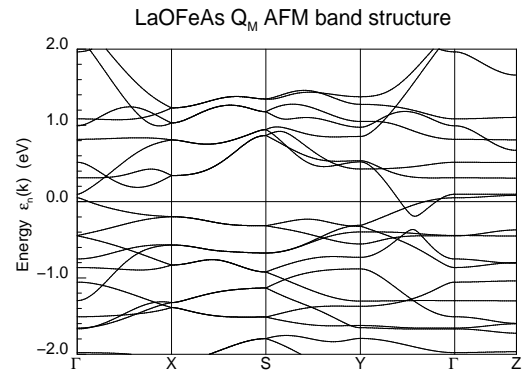}}}
\caption{Band structure of the $\vec Q_M$ AFM phase 
along high symmetry directions.  Note that two dispersive bands and one 
narrow band cross E$_F$ along $\Gamma$-Y, while only the one flatter
band crosses E$_F$ (very near $k$=0) along $\Gamma$-X.}
\label{bands}
\end{figure}

To prepare for studying the superconducting state, it is necessary first to
understand the normal state from which it emerges. 
Here we analyze the conducting $Q_M$ AFM phase, using results from two
all-electron, full potential codes Wien2k\cite{wien} and FPLO\cite{klaus1,klaus2}
using the generalized gradient approximation\cite{gga} (GGA)
functional.
We find the $Q_M$ phase to be energetically favored over the
$Q$=0 AFM phase by $\sim$75 meV/Fe, which itself lies 87 meV/Fe below the
nonmagnetic phase.
This energy difference is large enough that neither the $Q$=0 AFM,
nor
the nonmagnetic, phase will be thermally accessible at temperatures of 
interest. We neglect the antialignment of spins on the well separated
adjacent FeAs layers, which will have little effect on the electronic
and magnetic structure of a layer due to the weak interlayer hopping.
In either AFM 
phase, the Fe majority states are completely filled, thus
the moment is determined by the occupation of the minority states.
From the projected Fe $3d$ density of states (DOS) shown in Fig. \ref{dos}, 
the minority states are almost exactly
half-filled, giving 7.5 $3d$ electrons and thus an Fe state that is no
more than 0.5$e$ from neutral.
While the center of gravity of As $4p$ weight lies below that
of Fe $3p$ bands, there is strong mixing of these two
characters on both sides of E$_F$, and the As $4p$ states are certainly
unfilled.

\begin{figure}[tbp]
\rotatebox{-00}
{\resizebox{6.8cm}{8.0cm}{\includegraphics{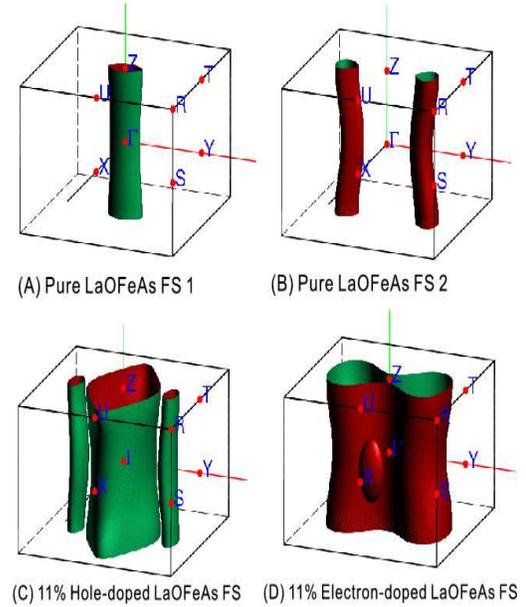}}}
\caption{(color online) Fermi surfaces of LaOFeAs. (A) and (B): the hole
cylinders and electron tubes of the stoichiometric $Q_M$ phase. (C) and
(D): hole- and electron-doped surfaces.
}
\label{FS}
\end{figure}

Notably,
the band structure and DOS is characterized by a pseudogap 
straddling E$_F$, closing
only in a small region along the $\Gamma$-Y line near $\Gamma$.  
Since the moments, and hence the exchange energies, of the two AFM phases
are very similar, the energy gain in the $Q_M$ phase can be ascribed
to the formation of the pseudogap. 
The system could
be considered as
{\it metallic} rather than semimetallic, in the sense that 
there are two dispersive bands crossing E$_F$ along $\Gamma$-Y.
One is 1.3 eV wide, comprised of Fe $d_{xy}$ + As $p_z$ character, 
the other of $d_{yz}$ character is 0.9 eV wide.  A third narrower (0.4 eV)
band of $3d_{x^2-y^2}$ character crosses E$_F$ near $\Gamma$.  The crossing 
of the dispersive bands along $\Gamma$-Y are such as to leave only 
two small distinct
2D Fermi surfaces, shown in Fig. \ref{FS}: 
an elliptical hole cylinder at $\Gamma$ containing $\sim$0.03 holes, and two
symmetrically placed near-circular
electron tubes midway along the $\Gamma$-Y axis.
In the sense that the 
Fermi surfaces are small,
the state is semimetallic.
The bands near E$_F$ 
have $k_z$ dispersion of no more than 25 meV.

The $d_{xz}, d_{yz}$
degeneracy is broken by the chains of aligned Fe spins in the $Q_M$
phase.  The rough
characterization for the minority Fe orbitals is that $d_{z^2}$ 
and $d_{x^2-y^2}$ states are partially filled, $d_{xy}$  and $d_{xz}$ 
states are empty,  the
$d_{yz}$ states are mostly filled but giving rise to the hole Fermi
surfaces. (Note that here the $x-y$ coordinate system is rotated by
45$^{\circ}$ from that usually used for the primitive cell, see Fig. 1.) 

A strikingly feature, crucial for accounting for observations, 
is that the DOS is (roughly)
particle-hole symmetric, as is the observed
superconducting behavior.  All bands near E$_F$
are essentially 2D, resulting in only slightly smeared
2D-like DOS discontinuities at the band
edges with structure elsewhere due to band crossings and non-parabolic 
regions of the bands.  The DOS has roughly a constant value of
0.25 states/(eV Fe spin) within 0.15-0.2 eV of E$_F$, with much flatter 
bands beyond.  The hole and electron effective 
in-plane masses, obtained from N(E) = $m^*/(\pi \hbar^2)$ for each pocket,
are $m^*_h=0.33, m^*_e = 0.25$.  An analogous band structure occurs in 
electron-doped HfNCl,\cite{weht} but there superconductivity appears
before the heavy bands are occupied.

\begin{figure}[tbp] \rotatebox{-90}
{\resizebox{6.0cm}{7.0cm}{\includegraphics{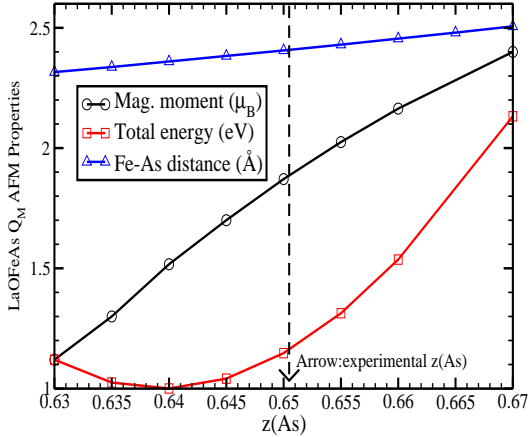}}}
\caption{(color online) The magnitude of the Fe magnetic moment, the
change in energy, and the Fe-As distance, as the As height $z_{As}$ is varied.}
\label{moment}
\end{figure}

There are somewhat conflicting indications of the possible importance
of electron-phonon coupling in this compound.\cite{boeri,eschrig}
Fig. \ref{moment} provides evidence of strong {\it magneto}phonon coupling:
increase of the As height which changes the Fe-As distance affects
the Fe moment at a rate of 6.8 $\mu_B$/\AA, indicating an unusually
large sensitivity to the Fe-As
separation.  Fig. \ref{moment} also reveals another important aspect:
LDA and GGA are almost 0.1 \AA~off in predicting the height of the As
layer relative to Fe, a discrepancy that is uncomfortable large.  Neglecting 
the Fe magnetism increases the discrepancy.

``Doping'' (change of charge in the FeAs layers) 
is observed to cause the N\'eel temperature
to decrease, and no magnetic order is apparent in superconducting
samples.
The effect of (rigid band or virtual crystal) doping on the 
$Q_M$ electronic structure, 
either by electrons
or holes, is to move $E_F$ into a region of heavier carriers, by roughly
a factor of 20 ($m^*_h \sim 6 \sim m^*_e$). About 0.1 carriers is 
sufficient to do this, which is just the amount of doping that results
in superconductivity.   The Fermi
surfaces evolve accordingly as shown in Fig. \ref{FS}: 
for electron doping the hole cylinder 
disappears, the electron tubes enlarge and merge; for hole doping the 
electron tubes decrease in size as the hole cylinder grows and distorts
into a diamond-shaped cross section.  

The spectrum of magnetic fluctuations  is an important property of
any AFM phase, and may bear strongly on the emergence of superconductivity.
We have calculated from linear response theory the  
exchange couplings $J_{ij}(q)$ for all pairs \{i,j\} within the unit cell, 
and by Fourier transform the real space
exchange couplings $J_{ij}(R)$,  for the transverse spin-wave
Hamiltonian\cite{spinwave1,spinwave2}
\begin{eqnarray}
H = -\sum_{<i,j>}J_{ij}\hat e_i \cdot \hat e_j;~~~
J_{ij}(R) =-\frac{d^2E[\{\theta\}]}{\partial \theta_i(0) \partial \theta_j(R)},
\end{eqnarray}
where $\theta_j(R)$ is the angle of the moment (with direction
$\hat e_j$) of the $j$-th spin in the
unit cell at $R$.  For the $Q_M$ AFM phase and experimental
structural parameters, the 1st and 2nd neighbor 
couplings are  
(distinguishing parallel and perpendicular spins)
\begin{eqnarray}
J_1^{\perp} = -550~{\rm K}; 
J_1^{\parallel} = +80~{\rm K}; 
J_2^{\perp} = -260~{\rm K}.
\end{eqnarray}  
For comparison, the nearest neighbor coupling\cite{spinwave1,spinwave2} 
in elemental FM Fe is $J_1 \approx 1850$ K, {\it i.e.} 3-4 times as strong.
The signs are all supportive of the actual ordering, there is no
frustration.
The factor-of-7 difference between the two 1st neighbor couplings reflects the
strong asymmetry between the $x$- and $y$-directions in the $Q_M$ phase, which
is also clear from the bands.  The sensitivity to the Fe-As distance is
strong: for $z$(As)=0.635, where the moment is decreased by 40\% (Fig.
\ref{moment}), the
couplings change by roughly a factor of two: $J_1^{\perp}$= -200 K,
$J_1^{\parallel}$ = +130 K, $J_2^{\perp}$ = -140 K.  
The interlayer exchange constants will be much smaller and, although
important for the (three dimensional) ordering, that coupling
should leave the spin-wave
spectrum nearly two-dimensional. 

We emphasize that these exchange couplings apply only to small rotations
of the moment (spin waves).
The $Q$=0 phase
couplings are different from those for the $Q_M$ phase; furthermore,
when FM alignment is enforced the magnetism disappears entirely.
The magnetic coupling is phase-dependent, largely itinerant,
and as mentioned above, it
is sensitive to the Fe-As distance. 

The electronic and magnetic structure, and the strength of magnetic 
coupling, in the reference state of the new iron arsenide superconductors
has been presented here, and the origin of the electron-hole symmetry 
of superconductivity has been clarified.  The dependence of the Fe moment
on the environment, and an unusually strong 
magnetophonon coupling, raises the possibility that magnetic fluctuations
are involved in pairing, but that it is longitudinal fluctuations that
are important here.

W.E.P. thanks Y. Tokura and P. B. Allen for discussions on this system.
We acknowledge support from NSF Grants DMR-0608283 and DMR-0606498
(S.Y.S.) and from DOE Grant DE-FG03-01ER45876 (W.E.P.).
S.L. acknowledges financial support from ANR PNANO Grant 
ANR-06-NANO-053-02 and ANR Grant ANR-BLAN07-1-186138.



\end{document}